\documentclass[
reprint,
amsmath,amssymb,
aps,
prl,
]{revtex4-1}
\usepackage{braket}
\usepackage{graphicx}
\usepackage{dcolumn}
\usepackage{bm}

\newcommand{\up}{\,\uparrow\,}
\newcommand{\dw}{\,\downarrow\,}

\begin{document}

\title{Collision of one-dimensional fermion clusters}

\author{Jun'ichi Ozaki}
\author{Masaki Tezuka}
\author{Norio Kawakami}

\affiliation{Department of Physics, Kyoto University, Kyoto, Japan 606-8502}

\date{\today}

\begin{abstract}
We study cluster-cluster collisions in one-dimensional Fermi systems with particular emphasis on the non-trivial quantum effects of the collision dynamics. 
We adopt the Fermi--Hubbard model and the time-dependent density matrix renormalization group method 
to simulate collision dynamics between two fermion clusters of different spin states with contact interaction. 
It is elucidated that the quantum effects become extremely strong with the interaction strength, 
leading to the transmittance much more enhanced than expected from semiclassical approximation. 
We propose a concise model based on one-to-one collisions, 
which unveils the origin of the quantum effects and also explains the overall properties of the simulation results clearly. 
Our concise model can quite widely describe the one-dimensional collision dynamics with contact interaction. Some potential applications, 
such as repeated collisions, are addressed. 
\end{abstract}


\maketitle

Recently non-equilibrium dynamics of cold atoms has attracted much attention, 
because cold atom systems are ideal as isolated quantum systems which can be experimentally designed \cite{review_cold_atom}. 
In cold atom systems, Feshbach resonance \cite{review_feshbach} has enabled experimentalists to modify strength and sign of interaction between atoms. 
Also, the dynamics of quantum quench \cite{exp_quench} has been explored by suddenly changing the trap potential and the interaction. 

Of these experiments, the dynamics of collision and mixing of two fermion clouds released from spin-dependent traps \cite{exp_spin_transport} motivates us 
to study collision dynamics between two fermion clusters in one-dimensional Fermi systems. 
For such one-dimensional collision dynamics, a well-defined semiclassical model exists \cite{c-c_collision_0}, 
so the fully quantum results and the semiclassical results can be theoretically compared. 
Thus the differences between the two results, which we call the quantum effects, can be calculated. 
The one-dimensional collision dynamics \cite{c-c_collision_0, c-c_collision_1, c-c_collision_2} has been simulated from the interest in that experiment. 
It has been shown that the sign of the interaction does not affect the dynamics, and that the non-trivial quantum effects are dominant when the interaction is strong.  

However, the origin of the quantum effects has not been clarified within our understanding. 
The complete analysis in the quantum effects of the collision dynamics is essential to an understanding of the quantum non-equilibrium dynamics, since the collision dynamics is one of the basic concepts of dynamics. 

In this study we simulate the cluster-cluster collision dynamics in one-dimensional spin-1/2 Fermi systems with contact interaction, 
and calculate the quantum many-body effects in the collision dynamics. 
Then we propose a quantum collision model which contains only the one-to-one collision parameters. 
This model, the independent collision model (ICM), reproduces well the simulation results, 
and fully elucidates the non-trivial quantum effects of the collision dynamics, which extremely enhance the transmittance at strong interaction. 
The ICM can quite widely explain the one-dimensional collision dynamics with contact interaction. 
We address the two applications of the ICM, interaction sign effects and repeated collisions.

\begin{figure}[htbp]
\begin{center}
\includegraphics[width=8.66truecm,clip]{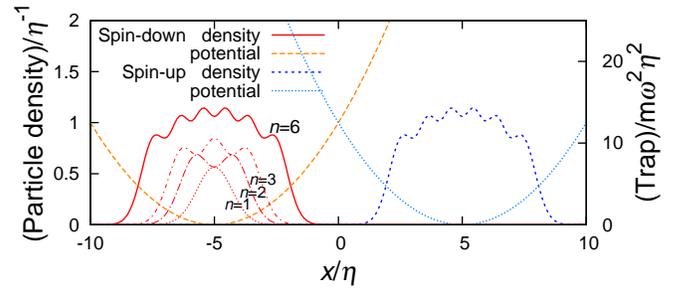}
\caption{(Color online) Initial particle density and trap potential at $n=6$, and initial spin-down density at $n=1$, $n=2$, $n=3$. }
\label{trap}
\end{center}
\end{figure}

We conduct simulations of collision dynamics, and calculate reflectance and transmittance of the clusters in one-dimensional spin-1/2 Fermi systems. 
Initially $n$ fermions per spin are trapped by spin-dependent potentials separately (Fig.\ref{trap}) ($n \leq 6$ for numerically exact results). 
Both trap potentials are harmonic, and they have the same shape but are spatially separated. 
The mass of a fermion is $m$, and the trap frequency of the harmonic potentials is $\omega$ (the oscillation cycle is $T = 2\pi/\omega$), and the interaction strength between the fermions is zero. 
So the typical width of the particle density tails is $\eta = \sqrt{\hbar/m\omega}$. 
We set the distance between the centers of the initial potentials as $2D = 10\eta \gg \eta$, 
so that the overlap between the two clusters is negligible. 
The center of the spin-down (up) trap potential is $x = -D = -5\eta$ ($x = D = 5\eta$).

At $t=0$ we suddenly change the trap potentials into a new shared potential, $V(x) = \frac{1}{2}m\omega^2 x^2$. 
Simultaneously we switch on the contact interaction between spin-up and spin-down fermions as $u \delta(x_d-x_u)$, where $x_d$ ($x_u$) is location of a spin-down (up) particle. 
Then the two clusters start moving towards each other without significantly changing their shapes, and they collide around $x = 0$ at $t = T/4$ with the average momentum $|p| = m\omega D$. 
Finally, depending on the interaction strength $u$, they are reflected to the initial location, or travel to the opposite location by $t=T/2$, 
and the particle density around $x=0$ is almost zero again. 
So the number of reflected particles $N_n^\mathrm{ref}(u)$ and the number of transmitting particles $N_n^\mathrm{tra}(u)$ are obtained 
by counting the spin-down (up) particle number in $x<0$ and $x>0$ ($x>0$ and $x<0$). 
We obtain the reflectance $R_n(u) = N_n^\mathrm{ref}(u)/n$ and the transmittance $T_n(u) = N_n^\mathrm{tra}(u)/n$ ($R_n(u) + T_n(u) = 1$). 
Clearly $R_n(0) = 0, T_n(0) = 1$ and $R_n(\infty) = 1, T_n(\infty) = 0$ because the system is one-dimensional.


We discretize the system to adopt the one-dimensional Fermi--Hubbard model, and apply the time-dependent density matrix renormalization group (t-DMRG) method \cite{t-DMRG, review_DMRG2, review_DMRG3} to simulate the dynamics. 
We take 199 sites numbered $-99, -98, \cdots , +98, +99$ at regular intervals; the site $-50$ ($+50$) is the initial location of the potential center for spin-down (up) atoms. 
The lattice constant is $\delta x = 2D/100 = 0.1\eta$, which is small enough to neglect the umklapp scattering. 
The trap potential of the site $i$ is $V_{i,\sigma}(t) = \frac{1}{2}m\omega^2(x_i - \sigma D)^2$ ($\sigma = -$ for spin-down, and $\sigma = +$ for spin-up). 
The discretized Hamiltonian is
\begin{eqnarray*}
H(t) &=& -\frac{\hbar^2}{2 m\delta x^2}\sum_{i,\sigma} (a^\dagger_{i,\sigma}a_{i+1,\sigma} + a^\dagger_{i+1,\sigma}a_{i,\sigma})\\ &+& \frac{u}{\delta x} \sum_{i} n_{i,+} n_{i,-} + \sum_{i,\sigma}V_{i,\sigma}(t)n_{i,\sigma} \;.
\end{eqnarray*}
We calculate the time evolution of this Hamiltonian starting from the ground state of the system by t-DMRG up to $t = T/2$, 
where the time step is $10^{-5}T$ and the maximum eigenvalue of the density matrix discarded is $\varepsilon < 10^{-12}$.
The simulation is conducted in the following range of parameters: the fermion number $n \leq 6$, the contact interaction strength $2^{-5} \leq u/u_c \leq 2^{5}$ ($u_c = 2\hbar p/m$). 

\begin{figure}[htbp]
\begin{center}
\includegraphics[width=8.66truecm,clip]{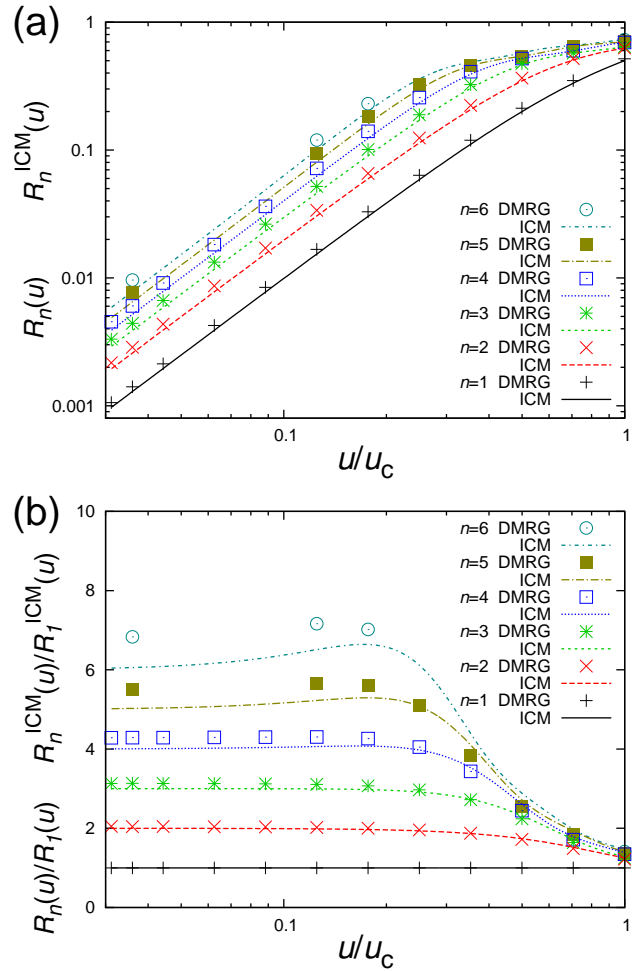}
\caption{(Color online) (a) Reflectance obtained by DMRG simulation $R_n(u)$ (dots) and that by ICM $R_n^\mathrm{ICM}(u)$ (lines) (log scale on $x$ and $y$ axes). 
(b) Reflectance ratio obtained by DMRG simulation $R_n(u)/R_1(u)$ (dots) and that by ICM $R_n^\mathrm{ICM}(u)/R_1^\mathrm{ICM}(u)$ (lines) (log scale on $x$ axis). }
\label{result_small}
\end{center}
\end{figure}

\begin{figure}[htbp]
\begin{center}
\includegraphics[width=8.66truecm,clip]{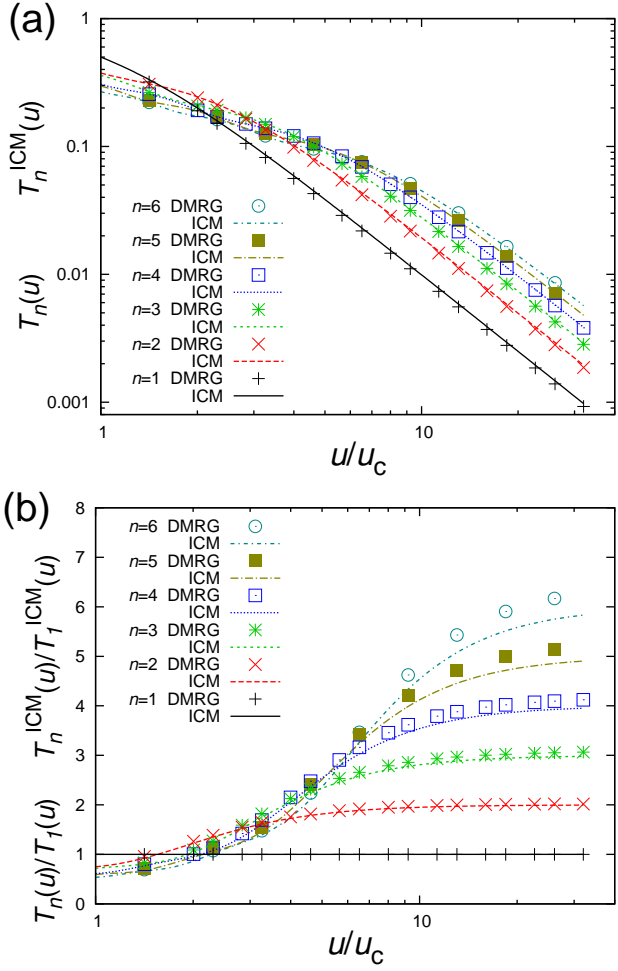}
\caption{(Color online) (a) Transmittance obtained by DMRG simulation $T_n(u)$ (dots)  and that by ICM $T_n^\mathrm{ICM}(u)$ (lines) (log scale on $x$ and $y$ axes). 
(b) Transmittance ratio obtained by DMRG simulation $T_n(u)/T_1(u)$ (dots)  and that by ICM $T_n^\mathrm{ICM}(u)/T_1^\mathrm{ICM}(u)$ (lines) (log scale on $x$ axis). }
\label{result_large}
\end{center}
\end{figure}

Figure \ref{result_small}(a) shows the reflectance $R_n(u)$ obtained by the DMRG simulation for $2^{-5} \leq u/u_c \leq 1$. 
The figure implies $R_n(u) \propto u^2$ in the small $u$ limit. 
We also plot the ratio $R_n(u)/R_1(u)$ in Fig.\ref{result_small}(b) to evaluate the many-body effects. 
It is observed that $R_n(u) \simeq n R_1(u)$ is approached in the limit of $u \rightarrow 0$. 
In this limit, almost all particles transmit, and a particle collides with the $n$ particles of different spin until it reaches the opposite side. 
So in the semiclassical picture, the reflectance is $nR_1(u)$ because the number of the reflected particles is approximately $n^2 R_1(u)$. 
Therefore there is no quantum effect in this limit \cite{c-c_collision_0}. 
On the other hand, Fig.\ref{result_large}(a) shows the transmittance $T_n(u)$ from the DMRG simulation for $1 \leq u/u_c \leq 2^{5}$.  
The figure shows $T_n(u) \propto u^{-2}$ in the large $u$ limit. 
We plot the ratio $T_n(u)/T_1(u)$ in Fig.\ref{result_large}(b), which illustrates $T_n(u) \simeq n T_1(u)$ as $u \rightarrow \infty$. 
In this limit, since almost all particles are reflected, in most cases a particle collides with another particle of different spin just once during a cluster-cluster collision. 
So in the semiclassical case, the number of the transmitting particles is almost $n T_1(u)$, so the transmittance is $T_1(u)$. 
Therefore there are strong quantum effects in this limit \cite{c-c_collision_0}, and the transmittance in the quantum case is $n$ times larger than in the semiclassical case.

We propose the independent collision model (ICM) to explain the simulation results. 
We start from the simplest case, the one-to-one collision dynamics, which can be easily calculated. 
The initial wavefunction is $\ket{\dw\up}$, which means that a spin-down (up) particle is in the left (right). 
When the spin-down particle with momentum $p$ and the spin-up particle with momentum $-p$ collide, 
$\ket{\dw\up}$ splits into $\rho\ket{\dw\up}$ (reflection term) $ + \tau\ket{\up\dw}$ (transmission term) and the momenta are reversed, in which 
$\rho = u/(i u_c - u)$, $\tau = i u_c/(i u_c - u)$.
The theoretical reflectance $R_1^\mathrm{ICM}(u)$ and transmittance $T_1^\mathrm{ICM}(u)$ are calculated 
by the coefficients of $\ket{\dw\up}$ and $\ket{\up\dw}$ as
\begin{eqnarray*}
R_1^\mathrm{ICM}(u) = |\rho|^2 = \frac{u^2}{u_c^2 + u^2}\;,\;\;\; T_1^\mathrm{ICM}(u) = |\tau|^2 = \frac{u_c^2}{u_c^2 + u^2}\; .
\end{eqnarray*}

Next we regard the multi-particle cases as a series of one-to-one collisions. 
In this system, the initial wavefunction can be expressed as a Hartree-Fock form. 
We set up the single-particle wavefunctions of spin-down particles as $\varphi_{-1}(x), \varphi_{-2}(x), \cdots, \varphi_{-n}(x)$, 
and those of spin-up particles as $\varphi_{1}(x), \varphi_{2}(x), \cdots, \varphi_{n}(x)$. 
So the initial wavefunction of the whole system is 
$\psi = (n!)^{-1}\left| \varphi_{-i}(x_{-j}) \right|_{\dw} \left| \varphi_{i}(x_{j}) \right|_{\up}$, 
where the variables in the determinant $\big| \cdot \big|_{\dw}$ ( $\big| \cdot \big|_{\up}$ ) belong to spin-down (up). 
We can take these single-particle wavefunctions not only as the eigenstates of the initial Hamiltonian, 
but also as the localized wavefunctions created by a unitary basis transformation. 
It is convenient to use the localized single-particle wavefunctions to analyze the cluster-cluster collision. 
The important property of the contact interaction is that there is no difference of the shape of the single-particle wavefunction before and after a one-to-one collision
but the whole wavefunction splits into the unchanged part (transmission term) and a spin-flipped part (reflection term). 
Therefore, if the $n^2$ one-to-one collisions occur independently, the time evolution of the system is described by 
(i) the time evolution of the single-particle wavefunctions and 
(ii) the splits of wavefunctions at the one-to-one collisions. 
The time evolution (i) occurs in each $\varphi_{\pm i}(x)$ and the form of $\varphi_{\pm i}(x)$ is changed, 
but this evolution does not change the expression of $\psi = (n!)^{-1}\left| \varphi_{-i}(x_{-j}) \right|_{\dw} \left| \varphi_{i}(x_{j}) \right|_{\up}$. 
On the other hand, at the time evolution (ii), the wavefunction splits into the transmission term and the reflection term. 
When the particle $a$ and the particle $b$ collide, $\psi$ splits into $\tau \psi + \rho F^{a}_{b}\psi$ 
($F^{a}_{b}$ is the flip operator of $\varphi_{a}$ and $\varphi_{b}$). 
Even when $a$ and $b$ have the same spin, that relation holds because $\tau + \rho F^{a}_{b} = 1$ for the particles of the same spin.

\begin{figure}[htbp]
\begin{center}
\includegraphics[width=6truecm,clip]{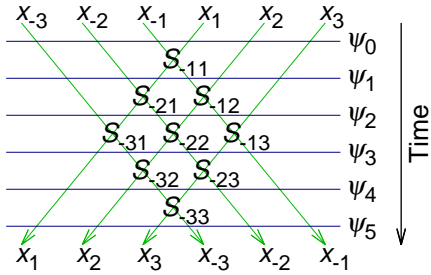}
\caption{(Color online) Collision order in ICM at $n=3$. $S_{-i j}$ is the one-to-one collision between $\varphi_{-i}$ and $\varphi_{j}$, and $\psi_k$ is the wavefunctions after the $k$th set of collisions. Simultaneous collisions are commutative. }
\label{model}
\end{center}
\end{figure}

We assume that $\varphi_{i}(x)$ is localized at $x_i$ ($x_{-n} < \cdots < x_{-1}$ and $x_{1} < \cdots < x_{n}$) and the time evolution (i) is expressed as the motion of $x_i$, 
and that $\varphi_{i}$ independently collides only with the wavefunctions of the reversed momentum at the same location. 
This assumption, the independent ordered collision, is supported numerically later. 
Since the movement of $x_i$ is similar to the case of the classical dynamics, the ordering of the collisions is the same (Fig. \ref{model}, which is discussed later). 
The initial wavefunction of the whole system is expressed as $\psi = \ket{\downarrow^{-n} \cdots \downarrow^{-1} \uparrow^{1} \cdots \uparrow^{n}}$, which means that 
there are spin-down particles at $x_{i}$ ($i < 0$) and spin-up at $x_{j}$ ($j > 0$) and 
the ordering of wavefunction in Slater determinant is adjusted to the one-dimensional ordering of $x_i$. 
The first one-to-one collision occurs between $x_{-1}$ and $x_1$.
So the wavefunctions splits into 
\begin{eqnarray*}
\tau \ket{\cdots \downarrow^{-2} \downarrow^{-1} \uparrow^{1} \uparrow^{2} \cdots} + \rho \ket{\cdots \downarrow^{-2} \uparrow^{-1} \downarrow^{1} \uparrow^{2} \cdots}\;,
\end{eqnarray*}
but the order of $x_{-1}$ and $x_{1}$ is reversed, so the expression
\begin{eqnarray*}
\rho \ket{\cdots \downarrow^{-2} \downarrow^{1} \uparrow^{-1} \uparrow^{2} \cdots} + \tau \ket{\cdots \downarrow^{-2} \uparrow^{1} \downarrow^{-1} \uparrow^{2} \cdots}\;
\end{eqnarray*}
corresponds to the spatial spin distribution. 
Simultaneous collisions are commutative, so we can assume that one of them occurs earlier. 
The next collision is assumed to be between $x_{-2}$ and $x_1$ (simultaneously $x_{-1}$ and $x_2$), then $\ket{\cdots \downarrow^{-2} \downarrow^{1} \uparrow^{-1} \uparrow^{2} \cdots}$ is not changed, 
but for the sign of Slater determinant, the spatial spin distribution expression is
\begin{eqnarray*}
\ket{\cdots \downarrow^{-2} \downarrow^{1} \uparrow^{-1} \uparrow^{2} \cdots} = -\ket{\cdots \downarrow^{1} \downarrow^{-2} \uparrow^{-1} \uparrow^{2} \cdots}\;.
\end{eqnarray*}
By omitting the ordering of $x_i$ from this expression, 
we obtain the simplified expression of the system wavefunction. 
It is possible to calculate the outcome of the cluster-cluster collision by calculating the system wavefunction after $n^2$ one-to-one collisions between all possible combinations of $x_{i}$ ($i < 0$) and $x_{j}$ ($j > 0$), 
because the single-particle wavefunctions evolve as that of free particles, and they collide $n$ times during the half cycle of the oscillation. 

Using the simplified expression, the cluster-cluster collision is calculated in the following process.
The initial wavefunction is simplified as $\ket{\dw\cdots\dw\up\cdots\up}$ with $n$ $\dw$'s and $n$ $\up$'s. 
The cluster-cluster collision is described by a series of one-to-one collisions, and 
(i) at the collision between $\dw$ and $\up$, the wavefunction splits into reflection term (amplitude $\rho$) and transmission term (amplitude $\tau$), 
(ii) at the collision between two particles of the same spin, the wavefunction is multiplied by $-1$. 
The ordering of the one-to-one collisions is the same as in the classical dynamics, shown in Fig. \ref{model}. 
The first collision occurs at the center of the system, and the second collisions occur at the locations next to the center ($n\geq 2$). 
The third collisions arise at the center and at the two locations away from the center ($n\geq 3$). 
Finally all the particles have the reversed momenta, and then collision dynamics finishes. 
The external parameters of the ICM are only $\rho$ and $\tau$, which are of the one-to-one collision. 
Since the calculations for large $n$ are complicated, we numerically compute the ICM reflectance $R_n^\mathrm{ICM}(u)$ and the ICM transmittance $T_n^\mathrm{ICM}(u)$.

We plot the reflectance obtained by the ICM $R_n^\mathrm{ICM}(u)$ in Fig.\ref{result_small}(a) and the ratio $R_n^\mathrm{ICM}(u) / R_1^\mathrm{ICM}(u)$ in Fig.\ref{result_small}(b) for $2^{-5} \leq u/u_c \leq 1$. 
The figure exhibits that $R_n^\mathrm{ICM}(u)$ agrees fairly well with $R_n(u)$ (the DMRG simulation loses accuracy in the limit of $u\rightarrow 0$), 
and the ICM is valid in the region. 
We discuss the asymptotic behavior of $R_n^\mathrm{ICM}(u)$ as $u\rightarrow 0$. 
In this limit, the dominant terms of $R_n^\mathrm{ICM}(u)$ are the coefficients of single-reflection wavefunctions (wavefunctions after only one reflection), 
since almost all particles transmit in this limit. 
The coefficients of the single-reflection wavefunctions are $\rho\tau^{n-1}$, and the number of the single-reflection wavefunctions is $n^2$, 
so $R_n^\mathrm{ICM}(u) \simeq n^2|\rho\tau^{n-1}|^2/n \simeq n(u/u_c)^2$ is the asymptotic form. 
This analytical calculation is free from the interference effects, so the absence of quantum effects in the limit of $u\rightarrow 0$ is supported by the ICM. 
Fig.\ref{result_large} also shows the ICM result $T_n^\mathrm{ICM}(u)$ (a) and the ratio $T_n^\mathrm{ICM}(u) / T_1^\mathrm{ICM}(u)$ (b) for $1 \leq u/u_c \leq 2^{5}$. 
The figure shows that $T_n^\mathrm{ICM}(u)$ is consistent with $T_n(u)$ (the difference comes from the momentum distribution and the uncertainty in the location of the collision), 
so the ICM is correct in the region too. 
The asymptotic behavior of $T_n^\mathrm{ICM}(u)$ in the limit of $u\rightarrow \infty$ is dominated by the coefficient of the component with a single transmission, $\ket{\dw\cdots\dw\up\dw\up\cdots\up}$. 
The coefficient is $\rho^{n-1}\tau(1 + \rho^2 + \cdots + \rho^{2n-2})$, so in the limit of $u \rightarrow \infty$ for fixed $n$, 
$T_n^\mathrm{ICM}(u) \simeq |\rho^{n-1}\tau(1 + \rho^2 + \cdots + \rho^{2n-2})|^2/n \simeq n(u/u_c)^{-2}$ is the asymptotic form. 
Different from the small $u$ case, this term contains the interference effect. 
In the semiclassical picture, in which all the horizontally aligned segments of the particle trajectories in Fig.\ref{model} are independently spin-down or spin-up at some probability determined by the previous stage, 
the interference vanishes and the transmittance becomes $|\rho^{n-1}\tau|^2(|1|^2 + |\rho^2|^2 + \cdots + |\rho^{2n-2}|^2)/n \simeq (u/u_c)^{-2}$. 
Thus the quantum transmittance is $n$ times larger than the classical transmittance, 
so the ICM unveils the quantum effects in the large $u$ limit.

\begin{figure}[htbp]
\begin{center}
\includegraphics[width=8.66truecm,clip]{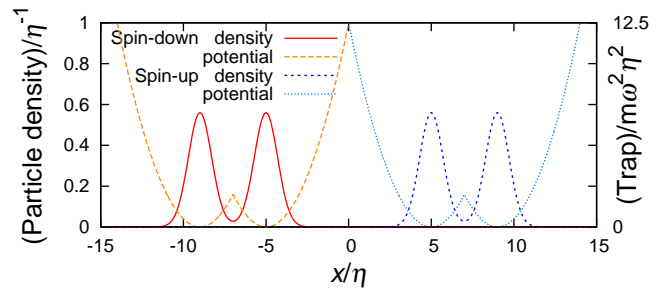}
\caption{(Color online) Initial particle density and trap potential at $n = 2$ and $d = 4\eta$. }
\label{trap_two}
\end{center}
\end{figure}

The ICM is consistent with the DMRG simulation, but we have assumed the independent ordered collision to derive the ICM. 
To support this assumption, we have also studied the following system. 
Initially $n$ fermions per spin are trapped separately, and the parameters of the particles are the same as those in the DMRG simulation of cluster-cluster collision (Fig.\ref{trap_two}). 
However, the trap potentials of both spins are not harmonic, but the combination of two harmonic potentials and the distance between these harmonic potentials is $d$. 
The harmonic centers of the spin-down (up) particles are at $x = -D, -D-d$ ($x = +D, +D+d$) ($D = 5\eta$). 

At $t=0$ we suddenly give a momentum $p$ to the spin-down particles and a momentum $-p$ to the spin-up particles ($p = m\omega D$), 
and simultaneously we switch off the trap potentials and set the contact interaction between spin-up and spin-down fermions as $u \delta(x_1-x_2)$. 
The two clusters start moving at the velocity $|v| = p/m$ and they collide at $x = 0$, then we calculate the reflectance $R^\mathrm{dist}_n(u,d)$. 
Except for the difference in the distribution of momentum or the decay of wavefunction, $R^\mathrm{dist}_n(u,0)$ should reproduce $R_n(u)$.
To simulate the system, we use the same method under the same condition as in the DMRG simulation above, except the following conditions.  
We calculate the time evolution up to $t = (2D+d)/(v-2\sqrt{m\omega\hbar}/m)$ for $n = 2$, $u/u_c  = 0.5, 1.0, 2.0$ and $0 \leq d \leq 4\eta$ ($\sqrt{m\omega\hbar}$ is the initial momentum variation). 
We take $299$ sites numbered $-149, -148, \cdots , +148, +149$ at regular intervals, but the lattice constant $\delta x = 0.1\eta$ is the same. 
The result of the simulation is that $|R^\mathrm{dist}_n(u,d) - R^\mathrm{dist}_n(u,0)|$ is under $1.5 \%$ for all $u$, so we conclude that the initial particle distribution has little effect on collision dynamics. 
Thus if we calculate the cluster-cluster collision, we can simplify the system by assuming that initially the fermions are independently localized and independently collide. 
In the simplified dynamics, the wavefunction of the system splits after every independent collision, and the ordering of the independent collisions is the same as in the classical dynamics. 
Therefore the assumption of the independent ordered collision is numerically supported. 




As an application of the ICM, we discuss the effects of the interaction sign in the cluster-cluster collision. 
In one-dimensional collision dynamics, the sign of the contact interaction $u$ does not affect the dynamics \cite{c-c_collision_0, c-c_collision_1}, 
and this property can be explained by the ICM. The imaginary unit $i$ is contained only in $\rho$ and $\tau$ in the ICM.
Since physical quantities such as particle density are real, the values of these quantities are not changed if we substitute $-i$ for $i$. 
After the substitution, $\rho(u)$ and $\tau(u)$ become $\rho(-u)$ and $\tau(-u)$. 
This implies that the physical quantities do not depend on the sign of the interaction. 

Another application of the ICM is the repeated cluster-cluster collision under the harmonic potential, 
in which the $k$th collision occurs at $(k-1)T/2 < t < kT/2$. 
The dynamics of the multiple cluster-cluster collisions was simulated \cite{c-c_collision_2}; the motion of center of mass is a linear function of $t$ in a region of strong interaction. 
We simulate this multiple cluster-cluster collision by the ICM, and calculate the wavefunction after the $k$th collision by using the final state after the $(k-1)$th collision as the initial state. 
We use their system parameters \cite{c-c_collision_2}, and assume the location of the localized wavefunctions $x_{-i} = -D - (i-(N+1)/2)\eta$ and $x_{i} = D + (i-(N+1)/2)\eta$ ($i > 0$). 
Thus we obtain the results that agree very well with their Fig. 3(b). 
Therefore the ICM is also useful for simulating the dynamics of the multiple cluster-cluster collisions.



In summary, using the time-dependent density matrix renormalization group method and the Fermi--Hubbard model, 
we have calculated the collision dynamics between two fermion clusters with the contact interaction. 
We have introduced the independent collision model for cluster-cluster collision, and theoretically checked the validity of our model 
by numerically showing that the initial distribution of particles does not affect the collision dynamics. 
Our model has reproduced the simulation results, and explained the large enhancement in the transmittance at strong interaction. 
Furthermore, we have demonstrated its potential applications to the interaction sign effect and the repeated collision dynamics.

This work was partially supported by the Grant-in-Aid for the Global COE Program ``The Next Generation of Physics, Spun from Universality and Emergence'' from MEXT of Japan. 
N. K. is supported by KAKENHI (Nos. 21740232, 20104010) and JSPS through its FIRST Program. 
J. O. is supported by a JSPS Fellowship for Young Scientists.







\end{document}